\documentclass{article}

\usepackage[compress]{cite}
\usepackage{amssymb}   

\usepackage{graphicx}  
\usepackage{amsmath}   
\usepackage{bm} 

\usepackage{mathrsfs}
\usepackage{color}

\def\eq#1{{Eq.~(\ref{#1})}}

\def\dqg{density of states of the mesoscopic spacetime}

\title{The Kinetic Theory of the Mesoscopic Spacetime\footnote{\textbf{This Essay received the Fourth Award in the Gravity Research Foundation Essay Contest 2018.}}}

\author{T. Padmanabhan\\
{\small IUCAA, Pune University Campus,
  Ganeshkhind, Pune - 411 007, India.}\\
\textit{email: paddy@iucaa.in}\\
}
\date{ }

\begin{document}

\maketitle

\begin{abstract}
At the mesoscopic scales ---  which interpolate between the macroscopic, classical, geometry and the microscopic, quantum, structure of spacetime --- one can identify the density of states of the  geometry which  arises from the existence of a zero-point length in the spacetime. This spacetime discreteness  also associates an internal degree of freedom with each event, in the form of a fluctuating vector of constant norm.
The equilibrium state, corresponding to the extremum of  the total density of states of geometry plus matter, leads precisely to Einstein's equations. In fact, the field equation can now be reinterpreted   as a zero-heat-dissipation principle. The analysis of fluctuations around the equilibrium state (which is described by Einstein's equations), will provide new insights about quantum gravity.
\end{abstract}

\noindent\textit{\textbf{The magic of kinetic theory: Counting the continuum:}}
The kinetic theory of a normal fluid is based%
\footnote{I use the signature is $(-,+,+,+)$ and the natural units with $c=1, \hbar =1$ and will set $\kappa = 8\pi G =8\pi L_P^2$ where $L_P$ is the Planck length $(G\hbar/c^3)^{1/2}=G^{1/2}$ in natural units. The  Latin letters $i, j$ etc. range over spacetime indices and the Greek letters $\alpha, \beta$ etc. range over the spatial indices. I will just write $x$ for $x^i$ etc., suppressing the index, when no confusion is likely to arise.}
on the  distribution function $f(x^i, p_j)$,  which \textit{counts}  the number of  degrees of freedom  $dN = f(x^i, p_j)d^3xd^3p$   per unit phase space volume $d^3xd^3p$ (with the constraint $p^2=m^2$ making the phase space  six-dimensional).
 This description, in terms of a distribution function $f$ --- which we could equivalently think of  as  the \textit{density of states} in phase space --- is remarkable because it achieves the impossible! It allows us to use the continuum language and, at the same time, take into account the discrete nature of the fluid. 
 The key new feature which enables this description is the  ``internal'' variable $p^\mu$ which   describes atoms (or molecules) with different \textit{microscopic} momenta co-existing at the same event $x^i$. 
 
 In an identical manner, we can introduce a function $\rho_g (x^i,\phi_A)$ capabale of  describing the \dqg. Here, $\phi_A$ (with $A=1,2,3,...$) denotes possible internal degrees of freedom (which are analogous to the momentum $p_i$ in the distribution function for the molecules of a fluid) that  exist as fluctuating \textit{internal} variables at \textit{each} event $x^i$. Their behaviour,  at any event $x^i$, is described by a probability distribution $P(\phi_A,x^i)$, the form of which depends on the microscopic quantum state of the spacetime. 
 
 There is a natural way of defining $\rho_g (x^i,\phi_A)$, if we introduce discreteness into the spacetime through a zero-point length. Remarkably enough, this procedure \textit{also} allows us to  identify  the internal variable $\phi_A$ as a four-vector $n^a$ with constant norm,  which can be thought of  as a microscopic, fluctuating, quantum variable at each event $x^i$. 
 Once we have determined the form of $\rho_g (x^i,n_a)\equiv\exp S_g$ and  the the corresponding density of states for matter,  $\rho_m \equiv\exp S_m$,  the equilibrium state is obtained by extremising  
 $\rho_g \rho_m =\exp[S_g+S_m]$. This, as we will see, leads to Einstein's equations, along with an elegant interpretation!

 \noindent 
\textit{\textbf{Spacetime events have finite areas but zero volumes:}} Let us start by determining the form of $\rho_g$. The two primitive geometrical constructs in any spacetime  are the area and the volume. It is, therefore,  natural to assume that the \dqg, $\rho_g(\mathcal{P})$,  at an event $\mathcal{P}$, is some function $F$ of either the area $A(\mathcal{P})$ or the volume $V(\mathcal{P})$  which we can ``associate with'' the event $\mathcal{P}$. Moreover, the total degrees of freedom, $\rho_g(\mathcal{P})\rho_g(\mathcal{Q})$, associated with two events $\mathcal{P}$ and $\mathcal{Q}$ are multiplicative
 while the primitive area/volume elements are additive. Therefore this function $F$ should be an exponential. In terms of the area, for example, we have: 
\begin{equation}
 \ln 
 \left\{ 
 \begin{array}{c}
  \text{density of states of the}\\
  \text{quantum geometry at $\mathcal{P}$}
 \end{array}
 \right\}\quad
 \propto\quad
\left\{ 
 \begin{array}{c}
  \text{area ``associated with''}\\
  \text{the event $\mathcal{P}$}
 \end{array}
 \right\}
\end{equation}
That is,
$\ln\rho_g(\mathcal{P})\propto A(\mathcal{P})$.

We next need to give  meaning to the phrase, area (or volume) ``associated with'' the event $\mathcal{P}$. 
To do this, let us consider the Euclidean extension of a local neighbourhood around $\mathcal{P}$ and  all possible geodesics emanating from  $\mathcal{P}$. Let $\mathcal{S}(\mathcal{P},\sigma)$ be the surface formed by all the events, at a given geodesic distance $\sigma$ from an event $\mathcal{P}$, which we will call an \textit{equi-geodesic surface} around the event
$\mathcal{P}$. Let $A(\mathcal{S})$ be its area and  $V(\mathcal{S})$ be the volume enclosed by this surface. The limiting values of  $A(\mathcal{S})$ and  $V(\mathcal{S})$ when $\sigma\to0$ will provide a natural definition of the area and  volume 
``associated with'' the event $\mathcal{P}$. 

In standard Riemannian geometry --- which knows nothing about the discreteness of microscopic quantum spacetime --- both the area and volume will vanish in the limit of $\sigma\to0$, as to be expected.  But when we introduce the discreteness of the spacetime through the existence of a zero-point length, we  find that \cite{paperD} the area associated with an event  becomes nonzero but the volume will still remain zero. Interestingly enough, this approach will also introduce an arbitrary, constant norm, vector $n_a$ into the description. (The norm of this vector is unity in the Euclidean sector; the vector will map to a  null vector with zero norm in the Lorentzian sector. See Fig. 1 of \cite{dice} for more details). The area ``associated with'' an event will therefore be a fluctuating, indeterminate, variable depending on a quantum degree of freedom $n^a$. 
It turns out that, in terms of this internal, vector degree of freedom, the $\rho_g(x^i,n_a)$ is given by
\begin{equation}
\ln \rho_g \propto \left[1- \frac{L_P^2}{8\pi} R_{ab}(x) n^an^b\right] =\mu\left[1-\frac{L_P^2}{8\pi}R_{ab}n^an^b\right]
\label{rhogresult}
\end{equation}
where $\mu$ is a dimensionless proportionality constant. 
We see that the term involving $R_{ab}$  comes in with a \textit{minus sign} in \eq{rhogresult}; this  is crucial for the success of our programme and we have \textit{no} control over it! 

\noindent
\textbf{\textit{The mesoscopic relic from the discreteness of quantum spacetime:}}
We see that $\rho_g$ depends on the extra, internal degree of freedom, $n_a$ which could take all possible values (at a given $x^i$) except for the constraint that it should have  a constant norm. \textit{This quantity is a direct relic of the discrete nature of the spacetime.} This is analogous to the momentum $p_j$ which appears in the fluid distribution function $f(x^i,p_j)$ --- as a relic of the discrete nature of the fluid ---   and again takes all possible values (except for the constant norm constraint,  $p^2=m^2$) at a given $x^i$.  
 The fluctuations of $n^a$ will be governed by some  functional $P[n^i(x),x]$, which is the probability that the quantum geometry is described by a vector field  $n_a(x)$ at each $x$. The exact form of $P$, of course,  is unknown at present, in the absence of the full theory of quantum gravity; but, fortunately, we will need only two properties of this  probability distribution $P[n^i(x),x]$ for our purpose: (i) It preserves the norm of $n^i$, which is unity in the Euclidean sector and is zero  in the Lorentzian sector; i.e  $P[n^i(x),x]$ will have the structure $F[n(x),x]\delta(n^2-\epsilon)$ with $\epsilon=1$ in the Euclidean space and zero in the Lorentzian spacetime. (ii) The   average value of $n^a$ over the fluctuations (in a given quantum state of the geometry), at any given event, gives,%
 \footnote{This is analogous to the fact that average value of the microscopic momenta of fluid particles $\langle p^\mu\rangle =P^\mu(x^i)$ gives rise to the macroscopic, mean, momentum of the fluid.  We know that the null normal $\ell_a$ also defines the tangent vector to the null geodesic congruence on the null surface; in this sense, it is indeed the momentum of, say, the photons traveling along the null geodesics.} 
 in the Lorentzian spacetime, a null normal $\ell^a(x^i)$ to a patch of null surface; i.e., we have $\langle n^a\rangle = \ell^a(x^i)$. 
 Of course, different quantum states of the spacetime geometry will lead to  different probability functionals $P$ with different null normals $\ell_a (x^i)$ as their mean values; so,  the expectation value $\langle n_a\rangle $ actually leads to the \textit{set of all null normals} $\{\ell_a(x^i)\}$ at an event $x^i$ when we take into account all the quantum states of the geometry.
  We can also write, $\langle n_in_j\rangle =\ell_i\ell_j+\sigma_{ij}$ where the second term $\sigma_{ij}$ represents quantum gravitational corrections to the mean value, etc. 
Therefore, the mean value $\langle \ln\rho_g(x^i,n_a)\rangle $,
 in the continuum limit, will be:
\begin{equation}
\langle \ln\rho_g(x^i,n_a)\rangle \propto 
 1-\frac{1}{8\pi} L_{P}^{2} R_{ab}\ell^a\ell^b +....
\label{denast2}
\end{equation}
where we have not displayed the terms proportional to $R_{ab}\sigma^{ab}$ which are of higher order and are independent of $\ell_a(x)$.

\noindent
\textit{\textbf{Density of states for matter:}} 
 In the continuum limit, it is quite straightforward to show \cite{A19} --- using the concept of local Rindler horizons associated with the null surface--- that  the density of states for matter is:
\begin{equation}
\langle \ln \rho_m\rangle \propto S_m \propto  L_P^4 T_{ab} \ell^a \ell^b=L_P^4\mathcal{H}_m
\label{rhomresult}
\end{equation}
where  $\mathcal{H}_m$ also has the  interpretion as the heat density contributed by matter crossing a null patch. If $T^a_b$ is due to an ideal fluid, then $T_{ab} \ell^a \ell^b=\rho+P=Ts$ is indeed the heat (entropy) density, where the last equality follows from the Gibbs-Duhem relation.  \eq{rhomresult} is a generalization of this result to \textit{any} $T^a_b$  as perceived by a local Rindler observer very close to the horizon.

\noindent
\textit{\textbf{The equilibrium state of matter and geometry:}} Taking into account both the matter and the spacetime, the total number of degrees of freedom, in the continuum limit --- in a state characterized by the mean vector field $\ell_a(x)$ --- will become:
\begin{equation}
 \langle \Omega_{\rm tot}\rangle_\ell  =\ \prod_{x}\, \langle \rho_g\rangle  \langle \rho_m\rangle  
 = \exp \sum_x \left( \langle \ln \rho_g\rangle  + \langle \ln \rho_m\rangle \right)\equiv \exp [S_{\rm grav}(\ell)+S_{\rm m}(\ell)]
 \label{omtot}
\end{equation}
to the lowest order (i.e., when we ignore the fluctuations, and set $\ln\langle \rho\rangle \approx\langle \ln\rho\rangle $).
The  $\ell_a$ dependent part of the configurational entropy $S_{\rm tot}=S_{\rm grav}+S_{\rm m}$ is then given by the functional
\begin{equation}
S_{\rm tot}[\ell(x)]=
\int_{\mathcal{S}} d^3V_x\; \mu E^a_b \langle n_an^b\rangle =\int_{\mathcal{S}} d^3V_x\;\mu\left(T^a_b (x) - \frac{1}{\kappa} R^a_b (x)\right) \ell_a(x)\ell^b(x) + ....
\label{av1}
\end{equation}
where, in the continuum limit, we have replaced the sum over $x$ by the integration over the null surface $\mathcal{S}$ for which $\ell^a(x)$ is the normal, with the measure
$d^3 V_x=(d\lambda d^2 x \sqrt{\gamma}/L_P^3)$ and introduced the proportionality constant $\mu$. 

The   field equations of gravity can be obtained by extremizing the expression for $\langle \Omega_{\rm tot}\rangle_\ell$ (or, equivalently, the configurational entropy $S_{\rm tot}=S_{\rm grav}+S_{\rm m}$) over $\ell$ and demanding that the extremum condition holds for all $\ell_a$. Since different quantum states of geometry will lead to different $\ell_a(x)$ at the same event $x^i$, this requirement is equivalent to demanding the validity of the extremum condition for all quantum states of the geometry, which have sensible classical limit.
Demanding that \eq{av1}  is an extremum with respect   to the variation $\ell^a \to \ell^a + \delta \ell^a$, 
(subject to the constraint $\ell^2=$ constant)  for all $\ell_a$ leads to $R^{a}_{b} - \kappa T^a_{b} = f(x) \delta^a_{b}$. Taking the divergence of this equation and using  $\nabla_a T^a_b =0$ and 
$\nabla_a R^a_b = (1/2)\partial_b R$, we get $f(x) = (1/2)R +$ a constant, leading to Einstein's equations, with the cosmological constant arising as an integration constant. 
So, the classical limit of the spacetime makes perfect thermodynamic sense.

\noindent
 \textit{\textbf{Einstein's equation interpreted as a Zero-Dissipation-Principle:}}
 What does this result actually mean? \textit{Unlike in the standard Einstein's theory, we now have a simple physical interpretation for the field equation!}
This arises from the fact that the quantity 
\begin{equation}
\mathcal{H}_g\equiv - \frac{1}{8\pi L_P^2}R_{ab}\ell^a\ell^b
\label{defhg}
\end{equation}
which determines the density of states of the quantum geometry, also
has an interpretation as the 
gravitational heat density (i.e., heating rate per unit area) of the null surface to which $\ell_a$ is the normal. Its integral over the null surface, $Q_g$, turns out to be the gravitational contribution to the heat content of the null surface. 
This result arises because the term $R_{ab}\ell^a\ell^b$ is related to the concept of ``dissipation without dissipation'' \cite{sanvedtp} of the null surfaces. 

Let me describe briefly how this interesting interpretation comes about. 
Start with the standard description of a null surface by introducing, in addition to the normal $\ell_a$, the complementary null vector $k^a$ (with $k^a\ell_a=-1$). The 2-metric on the cross-section of the null surface will then be  $q_{ab}=g_{ab}+\ell_ak_b+k_a\ell_b$. Define the expansion $\theta\equiv\nabla_a\ell^a$ and shear $\sigma_{ab}\equiv \theta_{ab}-(1/2)q_{ab}\theta$ of the null surface with $\theta_{ab}=q^i_aq^j_b\nabla_i\ell_j$. (We will assume that the null congruence is affinely parametrized.) One can then show that (see e.g., \cite{A19}) the integral of $R_{ab}\ell^a\ell^b$ over a null surface is
\begin{equation}
Q_{g}=-\frac{1}{8\pi L_P^2}\int \sqrt{\gamma}\, d^2x \, d\lambda\, R^a_b \ell_a\ell^b
=\int \sqrt{\gamma}\, d^2x \, d\lambda\, \left[2\eta \sigma_{ab}\sigma^{ab}+\zeta\theta^2\right]
\label{Qtoty}
\end{equation}
where the integrand
$
 \mathcal{D}\equiv\left[2\eta \sigma_{ab}\sigma^{ab}+\zeta\theta^2\right]
$
is the standard expression for the viscous heat generation rate of a fluid which has the shear and bulk viscous coefficients \cite{A26,A27,membrane} given by
 $\eta=1/16\pi L_P^2,\zeta=-1/16\pi L_P^2$. So we see that the $\mathcal{H}_g$ in \eq{defhg} is just the heat density of gravity on a null surface.

\textit{Of course, it will be unacceptable for every null surface to exhibit heating or dissipation!} This disaster is avoided by the presence of matter \textit{which is needed} if $R_{ab}\neq0$. The contribution to the heating from the microscopic degrees of freedom of the spacetime is precisely cancelled out by the heating  by the matter, on-shell.
 In fact, this fact allows us to reinterpret the field equation, expressed in the form:
\begin{equation}
-\frac{1}{8\pi L_P^2}R_{ab}\ell^a\ell^b +  T_{ab}\ell^a\ell^b=\mathcal{H}_g+\mathcal{H}_m= 0  
\label{zerodisp}
\end{equation} 
as a ``zero-heat-dissipation'' principle. 

 \noindent
 \textit{\textbf{Beyond Einstein: The spacetime fluid and its kinetics:}}
 In the usual kinetic theory of a normal fluid, the distribution function $f(x^a,p_i)$ --- which counts the microscopic degrees of freedom --- depends not only on $x^a$ but also on the internal variable $p_i$, which is a fluctuating four-vector of constant norm. This internal variable is a relic of the discrete nature of the fluid,  viz. it arises due to the existence of atoms/molecules of matter. 
 
 Similarly, when we introduce the kinetic theory of the mesoscopic spacetime, counting the corresponding mesoscopic  degrees of freedom of geometry $\rho(x^i,n_a)$ (through a simple limiting procedure, which associates an  area with each event), we \textit{discover} that it depends not only on $x^i$ but also on a internal variable $n_a$. This internal variable, again,  is  a fluctuating four-vector of constant norm and arises as a relic of the discrete microscopic nature of the spacetime fluid.

This \textit{discovery} of the mesoscopic spacetime degree of freedom $n_a$ enables us to define the equilibrium state for matter and geometry, purely from combinatorics --- viz., by maximizing the total number of degrees of freedom. The mean value of $n^a$ gives a null vector field $\ell^a(x)$ in the semi-classical limit. Different quantum  states of the geometry will, therefore, lead to different $\ell^a$-s at a given event. The demand that the extremum condition should hold for all $\ell^a$ at any given event, which is equivalent to demanding that the extremum condition holds for all quantum geometries; in other words, a  semi-classical spacetime arises only when this condition holds for all quantum geometries which are relevant for such a spacetime.
This extremum condition, in turn, gives us the Einstein's equations! 
 
 The concept of equilibrium  acquires a direct physical meaning in the classical limit, in which we identify the mean value $\langle n_a\rangle$ of the fluctuating internal variable with a null vector field $\ell_a(x)$ in the spacetime. The equilibrium condition is then equivalent to the statement $\mathcal{H}_g+\mathcal{H}_m=0$, where $\mathcal{H}_g$ is the dissipational heat density of gravity and $\mathcal{H}_m$ is the corresponding quantity for the matter. Equilibrium tantamounts to zero-heat-dissipation, as one would  expect. 
 In fact,  the fluctuations away from the equilibrium are governed by the standard Boltzmann factor
 \begin{equation}
F 
=\exp-\frac{1}{T_P}\left(\frac{1}{8\pi L_P^2} R^a_b (x) -T^a_b (x) \right) n_an^b
\label{boltz}
\end{equation} 
 where $T_P$ is the Planck temperature. The extremum condition, which leads to the vanishing of the argument of the exponent ensures that these degrees of freedom are not excited at the lowest order. This translates, in the classical limit, to the zero-heat-dissipation we obtained earlier.
 
 So we can now provide  a strikingly simple physical meaning for the gravitational field equations. In the standard form, Einstein's equation   $G^a_b=(8\pi L_P^2) T^a_b$, equates apples to oranges, viz., geometry to matter. In our case, we relate the geometrical heat of dissipation, $R^{ab}n_an_b$ (arising due to the coupling of internal variable $n_a$ with geometry) and  heat of dissipation of matter, $T^{ab}n_an_b$ (arising due to the coupling of internal variable $n_a$ with matter). 
 
 This also suggests a possible way of addressing the question: What is the \textit{actual mechanism} by which $T^a_b$ generates $R^a_b$ ? In Einstein's theory this is just a hypothesis, in the form of the field equation. In our approach, both the  spacetime geometry and matter couples to the internal variable $n_a$ through the terms $R^a_b n_a n^b$ and $T^a_b n_a n^b$ respectively, thereby leading to an effective coupling between them. Einstein's equation is just an average, equilibrium condition and we will expect --- as in any statistical system involving large number of degrees of freedom --- fluctuations around this equilibrium. This opens up new, exciting vistas of exploration.
 
 \noindent
 \textit{\textbf{Acknowledgements:}} My research is partially supported by the J.C. Bose Fellowship of SERB, India.
  

\end{document}